\documentclass[12pt,a4]{article}

\usepackage{a4wide}
\usepackage{amsmath,amsfonts,amsthm,amssymb,cite,bm}
\usepackage[dvips]{graphicx}
\usepackage{times,txfonts,eucal}
\numberwithin{equation}{section}

\def\6{\partial}
\def\<{\langle}
\def\>{\rangle}
\def\C{\bm{\mathit{C}}}
\def\D{\bm{\mathit{D}}}
\def\E{\bm{\mathit{E}}}
\def\A{\bm{\mathit{A}}}
\def\dd{\bm{\mathit{d}}}
\def\ee{\bm{\mathit{e}}}
\def\d{\mathrm{d}}
\def\double>{\rangle\!\rangle}
\def\im{\mathrm{i}}
\def\e{\mathrm{e}}
\def\d{\mathrm{d}}

\begin{document}
\baselineskip 16pt
\parskip 8pt


\title{Two-Species Asymmetric Simple Exclusion Process
	with Open Boundaries}
\author{Masaru Uchiyama \footnote{\tt{uchiyama@monet.phys.s.u-tokyo.ac.jp}}\\ \\
\emph{Department of Physics, Graduate School of Science,}\\
\emph{University of Tokyo,}\\
\emph{Hongo 7-3-1, Bunkyo, Tokyo 113-0033, Japan.}}
\date{}

\maketitle


\begin{abstract}
We study the steady state of 
the two-species Asymmetric Simple Exclusion Process (ASEP) 
with open boundary conditions. The matrix product method works 
for the determination of the stationary probability distribution. 
Several physical quantities are calculated through an explicit 
representation for the matrix products. 
By making full use of the relation with the continuous big $q$-Hermite 
polynomials, we arrive at integral formulae for the partition function 
and the $n$-point functions. 
We examine the thermodynamic limit and find three phases: the low-density phase,
the high-density phase and the maximal current phase. 
\end{abstract}


\section{Introduction}

Nonequilibrium behaviors of physical systems are one of the most important 
theme of the statistical physics. 
Although the general theory is partially absent, 
a lot of efforts have been made to study several kinds of simplified models 
that possess nonequilibrium properties. 
One of the most studied model is the Asymmetric Simple Exclusion Process (ASEP). 
This is a model of many particles on a one-dimensional system. 
Due to an exclusion interaction, the model essentially belongs to a many-body system, 
however, the model is solvable and deep analysis can be made 
by means of the matrix product method. 
There are many variants for the ASEP. The two-species version of the ASEP 
exhibits remarkable behaviors in many ways, such as 
construction of a shock profile \cite{2ASEP-1,2ASEP-2,Derrida-rev} 
and spontaneous translation breaking \cite{Arndt1,Arndt2,Sasamoto00,Rajewsky}. 

One of the most interesting findings for the ASEP is the relation with 
the theory of the $q$-orthogonal polynomials \cite{Sasamoto99,Blythe00,Sasamoto00-1,
USW,UW05}. 
In this paper, we extend the relation for the two-species version of the ASEP 
with open boundary conditions. 
We will see that the continuous big $q$-Hermite polynomials play a central role 
for this model. 
By the help of the $q$-calculus, 
we calculate the bulk quantities and the $n$-point functions for the steady state
and obtain integral formulae for them. 
The integral formulae are useful for evaluation in the thermodynamic limit. 
We clarify the existence of the boundary-induced phase transitions. 

This paper is organized as follows. 
In Sec. \ref{sec:2asep_model} the model is introduced and 
the matrix product method is presented 
for the two-species ASEP with open boundaries. 
In Sec. \ref{sec:q} a preliminary for the $q$-calculus is given. 
Several formulae for the continuous big $q$-Hermite polynomials are 
presented. 
In Sec. \ref{sec:rep} an explicit representation for the matrices and vectors 
in the matrix product method is provided. This representation is the 
key for understanding the relation between the model and 
the continuous big $q$-Hermite polynomials. 
In Sec. \ref{sec:analysis} an application of the formulae in Sec. \ref{sec:q} 
is made to the calculation of the quantities for the steady state of the model. 
We obtain integral formulae for the partition function and the $n$-point 
functions. 
By taking the thermodynamic limit, phase transitions are appeared. 
A comparison with the result for the one-species ASEP is made. 
The last section is for conclusion.

\section{Model}
\label{sec:2asep_model}

\begin{figure}[tb]
\begin{center}
\includegraphics{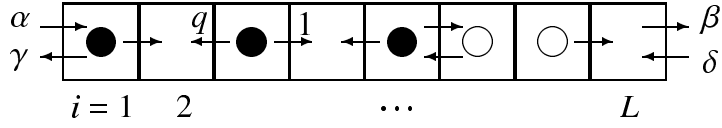}
\end{center}
\caption{The two-species ASEP with open boundaries.
	The black circle denotes the particle $2$ and 
	the white one denotes the particle $1$.
	Only the particle $2$ can get in or get out at the boundaries.
\label{fig:2asep}
}
\end{figure}

The ASEP is a one-dimensional stochastic process of many particles. 
Take a one-dimensional lattice of size $L$ with two boundaries. 
Consider two species of particles, denoted by $1$ and $2$ 
for the first- and second-class particle, respectively. Let $0$ denote 
an empty site. We assume an exclusion interaction of the particles such that 
there is at most one particle, $1$ or $2$, on each site. 
Particles hop in a random way on the lattice. 
The rates of hopping to the right and left nearest sites are fixed asymmetric, 
say $p_R$ and $p_L$. By scaling time, we put $p_R=1$ and $p_L=q$. 
That is, the dynamics of the particles in the bulk is 
\begin{align}
10\ \mathop{\rightleftarrows}^1_q\ 01, \qquad
20\ \mathop{\rightleftarrows}^1_q\ 02, \qquad
21\ \mathop{\rightleftarrows}^1_q\ 12.
\label{eq:exchange_ASEP2}
\end{align}
We mainly study the case $0\le q<1$ 
since the case $q>1$ is mapped by the parity transformation. 
The symmetric case $q=1$ is special. 
Since we are interested in the nonequilibrium behavior, we postpone the symmetric case. 
At the boundaries, the injection and removal of the particle $2$ are allowed. 
At the left (resp. right) boundary,
the injection rate is $\alpha$ (resp. $\delta$), 
and the removal rate is $\gamma$ (resp. $\beta$) if the target site is empty. 
We take $\alpha,\beta,\gamma,\delta$ positive. 
If we choose $\beta=\delta=0$ for example, the system will encounter 
jamming of particles. We do not consider such an extreme situation. 
We assume that the injection and removal of the particle $1$ are not allowed, 
and hence the number of the particle $1$ is fixed to $N$. 
The system is illustrated in Figure \ref{fig:2asep}. 
If the particle $1$ is absent in this system, we have the one-species ASEP. 
We remark that the dynamics of the particle $2$ in the bulk 
is irrespective of the presence of 
the particle $1$ (see Eq.(\ref{eq:exchange_ASEP2})), 
while the injection and removal of the particle $2$ at the boundaries 
can be affected by the particle $1$ 
due to the exclusion interaction between the particles $1$ and $2$.

Let $\tau_i=1,0$ and $\sigma_i=1,0$ be 
the particle numbers at site $i$ for the particles $2$ and $1$, respectively. 
The stochastic process is defined as the time-evolution of the probability 
distribution $P(\{\tau_i,\sigma_i\};t)$ with respect to the particle configuration 
$\{\tau_i, \sigma_i\}=(\tau_1,\cdots,\tau_L,\sigma_1,\cdots,\sigma_L)$. 
We use the bra-ket vector form in the basis of 
$\left\{\bigotimes_{i=1}^L |\tau_i,\sigma_i\double> 
\big|\tau_i=1,0, \sigma_i=1,0\right\}$. 
The master equation is described by 
\begin{align}
\frac{\d}{\d t}|P(\{\tau_i,\sigma_i\};t)\double>=
H|P(\{\tau_i,\sigma_i\};t)\double>.
\end{align}
The evolution operator is defined as
\begin{align}
H=\sum_{i=1}^{L-1} H_{ii+1}+H_1+H_L
\end{align}
with local operators 
\begin{align}
H_{ii+1}=
\left[
\begin{array}{ccc|ccc|ccc}
0&0&0&0&0&0&0&0&0\\
0&-1&0&q&0&0&0&0&0\\
0&0&-1&0&0&0&q&0&0\\ \hline
0&1&0&-q&0&0&0&0&0\\
0&0&0&0&0&0&0&0&0\\
0&0&0&0&0&-1&0&q&0\\ \hline
0&0&1&0&0&0&-q&0&0\\
0&0&0&0&0&1&0&-q&0\\
0&0&0&0&0&0&0&0&0
\end{array}
\right],
\end{align}
and 
\begin{align}
H_1=
\left[
\begin{array}{ccc}
-\gamma&0&\alpha\\
0&0&0\\
\gamma&0&-\alpha
\end{array}
\right],
\qquad
H_L=
\left[
\begin{array}{ccc}
-\beta&0&\delta\\
0&0&0\\
\beta&0&-\delta
\end{array}
\right].
\end{align}

We are mainly interested in the steady state of the system. 
The stationary probability distribution can be constructed 
by the matrix product method. 
We prepare the matrices $\D$, $\E$ and the vectors $\< W|$, $|V\>$ satisfying 
\begin{align}
H_{ii+1} \left(
\begin{array}{c}
\D\\ \A\\ \E
\end{array}
\right)_i\otimes\left(
\begin{array}{c}
\D\\ \A\\ \E
\end{array}
\right)_{i+1}=
\left(
\begin{array}{c}
\D\\ \A\\ \E
\end{array}
\right)_i\otimes\left(
\begin{array}{c}
\bar{\D}\\ \bar{\A}\\ \bar{\E}
\end{array}
\right)_{i+1}-\left(
\begin{array}{c}
\bar{\D}\\ \bar{\A}\\ \bar{\E}
\end{array}
\right)_i\otimes\left(
\begin{array}{c}
\D\\ \A\\ \E
\end{array}
\right)_{i+1},
\label{eq:MPA_rel_2asep}
\end{align}
\begin{align}
H_1 \< W|\left(
\begin{array}{c}
\D\\ \A\\ \E
\end{array}
\right)_1=
\< W|\left(
\begin{array}{c}
\bar{\D}\\ \bar{\A}\\ \bar{\E}
\end{array}
\right)_1,
\qquad
H_L \left(
\begin{array}{c}
\D\\ \A\\ \E
\end{array}
\right)_L|V\> =
-\left(
\begin{array}{c}
\bar{\D}\\ \bar{\A}\\ \bar{\E}
\end{array}
\right)_L|V\>
\end{align}
with some $\bar{\D}$, $\bar{\E}$, $\bar{\A}$.
The choice $\bar{\D}=I, \bar{\E}=-I, \bar{\A}=O$ is easy to treat.  
This gives the relations 
\begin{subequations}
\begin{align}
&\D\E-q\E\D=\D+\E,\label{eq:MPA-DE}\\
&\A\E-q\E\A=\A,\label{eq:MPA-AE}\\
&\D\A-q\A\D=\A,\label{eq:MPA-DA}\\
&\< W|(\alpha \E-\gamma \D)=\< W|,\label{eq:MPA-W}\\
&(\beta \D-\delta \E)|V\>=|V\>.\label{eq:MPA-V}
\end{align}
\label{eq:MPA}
\end{subequations}
Then, we have the steady state in a matrix product expression, 
\begin{align}
P\left(\{\tau_i,\sigma_i\}\right)=
\frac{1}{Z_{L,N}} 
\< W|\prod_{i=1}^{\underrightarrow{\ \ L\ \ }} 
(\tau_i \D+\sigma_i \A+(1-\tau_i)(1-\sigma_i)\E)|V\> ,
\end{align}
where $Z_{L,N}$ is the normalization constant. 
We often call the normalization as the partition function. 
It is convenient to consider the grand canonical ensemble. 
Introducing the fugacities $\xi$ and $\zeta$ for the particles $2$ and $1$, 
respectively, 
the steady state is generalized to 
\begin{align}
P\left(\{\tau_i,\sigma_i\};\xi,\zeta\right)=
\frac{1}{Z_{L}(\xi, \zeta)} 
\< W|\prod_{i=1}^{\underrightarrow{\ \ L\ \ }} 
(\tau_i \xi \D+\sigma_i \zeta \A+(1-\tau_i)(1-\sigma_i)\E)|V\> ,
\label{eq:MPA_2asep}
\end{align}
where the partition function is simply written as 
\begin{align}
Z_L(\xi,\zeta)=\< W|(\xi \D+\zeta \A+\E)^L|V\> . 
\end{align}
The case $\zeta=0$ gives the one-species ASEP. 
In fact, the $\xi\neq 1$ case is achieved just by modifying 
the boundary parameters and the fugacity $\zeta$, 
hence in the below 
we often consider the $\xi=1$ case and omit $\xi$ without any confusion. 
The canonical partition function is reproduced by an integral around the origin, 
\begin{align}
Z_{L,N}=\oint \frac{\d\zeta}{2\pi \im\zeta^{N+1}} Z_L(\xi=1, \zeta).
\end{align}

We can obtain physical quantities for the ASEP as follows. 
We denote the expectation value of $\mathcal{O}$ 
with respect to the steady state by $\<\mathcal{O}\>$. 
For the grand canonical ensemble, 
the particle densities and their variances are calculated as 
\begin{align}
&\rho_2=\<\tau_i\>=\frac{\xi}{L}\frac{\6}{\6\xi}\log Z_L(\xi,\zeta)\big|_{\xi=1},
\qquad \Delta\rho_2^2=
\left(\frac{\xi}{L}\frac{\6}{\6\xi}\right)^2\log Z_L(\xi,\zeta)\big|_{\xi=1},
\label{eq:rho_2}\\
&\rho_1=\<\sigma_i\>=\frac{\zeta}{L}\frac{\6}{\6\zeta}\log Z_L(\xi,\zeta),
\qquad\quad \Delta\rho_1^2=
\left(\frac{\zeta}{L}\frac{\6}{\6\zeta}\right)^2\log Z_L(\xi,\zeta).
\label{eq:rho_1}
\end{align}
Since the particle hopping is set asymmetric, there is a particle current even 
in the steady state. 
The particle current is calculated by the matrix product method as follows. 
We have the particle current of the particle $2$ in the grand canonical ensemble as 
\begin{align}
J_2&=\mathrm{Prob}\{\cdots 20\cdots\}-q\mathrm{Prob}\{\cdots 02\cdots\}
+\mathrm{Prob}\{\cdots 21\cdots\}-q\mathrm{Prob}\{\cdots 12\cdots\}\nonumber\\
&=\frac{1}{Z_L(\zeta)}\< W|\C^{i-1}(\D\E-q\E\D+\D\zeta \A-q\zeta \A\D)\C^{L-i-1}|V\> \nonumber\\
&=\frac{Z_{L-1}(\zeta)}{Z_L(\zeta)},
\label{eq:J_2-1}
\end{align}
where we have used (\ref{eq:MPA-DE}) and (\ref{eq:MPA-DA}). 
Similarly, in the canonical ensemble we have 
\begin{align}
J_2=\frac{Z_{L-1,N}}{Z_{L,N}}.
\label{eq:J_2-2}
\end{align}
We have the particle current of the particle $1$ in the grand canonical ensemble as
\begin{align}
J_1&=\mathrm{Prob}\{\cdots 10\cdots\}-q\mathrm{Prob}\{\cdots 01\cdots\}
-\mathrm{Prob}\{\cdots 21\cdots\}+q\mathrm{Prob}\{\cdots 12\cdots\}\nonumber\\
&=\frac{1}{Z_L}
\< W|\C^{i-1}(\zeta \A\E-q\E\zeta \A-\D\zeta \A+q\zeta \A\D)\C^{L-i-1}|V\> 
\nonumber\\
&=0.
\end{align}
Obviously, we have $J_1=0$ in the canonical ensemble. 


\section{Continuous big $q$-Hermite polynomials}
\label{sec:q}

The ASEP is closely related with the theory of the $q$-orthogonal polynomial. 
In the two-species model considered here, the continuous big $q$-Hermite 
polynomials play the central role. 
In this section, we give a preliminary for the $q$-calculus exploited in the analysis. 
For more information and summaries on the $q$-calculus and the $q$-orthogonal 
polynomials, please refer to \cite{Askey-Wilson,Gasper-Rahman,Koekoek}.
Note that formulae appeared in this section hold for $|q|<1$. 

We recall the multiple $q$-shifted factorial for $n=1,2,\cdots$: 
\begin{align}
(a_1,\cdots,a_s;q)_n=\prod_{r=1}^s\prod_{k=0}^{n-1} (1-a_r q^k).
\end{align}
For $n=0$, it is defined to be $1$. 
The basic hypergeometric function is defined as 
\begin{align}
{}_r\phi_s\left[
{{a_1,\cdots,a_r}\atop{b_1,\cdots,b_s}}\bigg|q;z\right]
=\sum_{k=0}^\infty \frac{(a_1,\cdots,a_r;q)_k}{(q,b_1,\cdots,b_s;q)_k}
\left((-)^k q^{\left(k\atop 2\right)}\right)^{1+s-r} z^k.
\end{align}
In terms of the basic hypergeometric function, 
the continuous big $q$-Hermite polynomial is defined as 
\begin{align}
H_n(x;\zeta'|q)&=\zeta'^{-n}{}_3\phi_2\left[
{{q^{-n},\zeta' \e^{\im\theta},\zeta' \e^{-\im\theta}}\atop{0,0}}
\bigg|q;q\right]\\
&=\sum_{k=0}^n \frac{(q;q)_n}{(q;q)_k(q;q)_{n-k}} (\zeta' \e^{\im\theta};q)_k
\e^{\im(n-2k)\theta}
\end{align}
with $x=\cos\theta$. 
Note that $H_n(x;\zeta'|q)$ is a $q$-orthogonal 
polynomial with respect to $x$ of degree $n$ 
as a descendant of the Askey-Wilson polynomial: 
\begin{align}
H_n(x;\zeta'|q)=P_n(x;\zeta',0,0,0|q).
\end{align}
The three-term recurrence relation is known as 
\begin{align}
H_{n+1}(x;\zeta'|q)+\zeta' q^n H_n(x;\zeta'|q)+(1-q^n) H_{n-1}(x;\zeta'|q)
=2xH_n(x;\zeta'|q)
\label{eq:rec-bigHermite}
\end{align}
with $H_{-1}=0$ and $H_0=1$. 
The orthogonality relation of $H_n(x;\zeta'|q)$ reads 
\begin{align}
\int_{0}^\pi \frac{\d\theta}{2\pi}
\frac{(q,\e^{2\im\theta},\e^{-2\im\theta};q)_\infty}
	{(\zeta' \e^{\im\theta},\zeta' \e^{-\im\theta};q)_\infty}
H_m(\cos\theta;\zeta'|q)H_n(\cos\theta;\zeta'|q)
=(q;q)_n\delta_{mn}.
\label{eq:ortho-bigHermite}
\end{align}
We introduce supplementary two-variable polynomials: 
\begin{align}
Q_n(x,y;t_1,t_2)=(t_1t_2xy;q)_n t_1^{-n} {}_3\phi_2\left[
{{q^{-n},t_1x,t_1y}\atop{t_1t_2xy,0}}\bigg| q,q\right] .
\end{align}
In particular, putting $t_1=\lambda$ and $t_2=0$ in $Q_n(x,y;t_1,t_2)$, we have
\begin{align}
F_n(x,y;\lambda)=\sum_{k=0}^n \frac{(q;q)_n}{(q;q)_k(q;q)_{n-k}}
(\lambda x;q)_k y^k x^{n-k},
\end{align}
and this polynomial satisfies the recurrence relation
\begin{align}
F_{n+1}(x,y;\lambda)+\lambda xyq^n F_n(x,y;\lambda)+(1-q^n)xy F_{n-1}(x,y;\lambda)
=(x+y)F_n(x,y;\lambda)
\end{align}
with $F_{-1}=0$ and $F_0=1$. 
Obviously, we have 
$H_n(\cos\theta;\zeta'|q)=F_n(\e^{\im\theta},\e^{-\im\theta};\zeta')$. 

The bilinear sum formula for $Q_n(x,y;t_1,t_2)$ holds as follows. 
When $t_1t_2xy=s_1s_2zw$, 
\begin{align}
&\sum_{n=0}^\infty
\frac{\tau^n}{(q,s_1s_2zw;q)_n} Q_n(x,y;t_1,t_2)Q_n(z,w;s_1,s_2)\nonumber\\
&=\frac{(\tau t_1xyz, \tau t_2xyz, \tau s_1xzw, \tau s_2xzw;q)_\infty}
{(\tau t_1t_2x^2yz, \tau xw, \tau yz, \tau xz;q)_\infty}\nonumber\\
&\qquad\times
{}_8W_7\left[ \tau t_1t_2x^2yz/q; t_1x, t_2x, s_1z, s_2z, \tau xz\big|
q,\tau yw\right] ,
\label{eq:sumQQ}
\end{align}
where a very-well-poised ${}_{r+1}\phi_r$ series is shortly written: 
\begin{align}
{}_{r+1}W_r(a_1; a_4,\cdots, a_{r+1}| q,z)=
{}_{r+1}\phi_r\left[
{{a_1,qa_1^{1/2},-qa_1^{1/2},a_4,\cdots,a_{r+1}}\atop{a_1^{1/2},-a_1^{1/2},
qa_1/a_4,\cdots,qa_1/a_{r+1}}}
\bigg| q,z\right].
\end{align}
If we put $x=\e^{\im\theta}$, $y=\e^{-\im\theta}$, $z=\e^{\im\phi}$, $w=\e^{-\im\phi}$, 
Eq. (\ref{eq:sumQQ}) is reduced to a bilinear sum for the Al-Salam-Chihara polynomial. 
The proof of (\ref{eq:sumQQ}) is in the same line as \cite{Ismail}.
We need specializations of (\ref{eq:sumQQ}). 
Setting $t_1=s_1=\zeta'$, $t_2=s_2= 0$ and 
$x=\e^{\im\theta}$, $y=\e^{-\im\theta}$, $z=\e^{\im\phi}$, $w=\e^{-\im\phi}$ gives
\begin{align}
&\sum_{n=0}^\infty \frac{\tau^n}{(q;q)_n}
H_n(\cos\theta;\zeta'|q)H_n(\cos\phi;\zeta'|q)
\nonumber\\
&=\frac{(\tau,\tau\zeta' \e^{\im\phi},\tau\zeta' \e^{-\im\phi};q)_\infty}
	{(\tau \e^{\im(\theta+\phi)},\tau \e^{\im(\theta-\phi)},
	\tau \e^{-\im(\theta+\phi)},\tau \e^{-\im(\theta-\phi)};q)_\infty}
{}_3\phi_2\left[
{{\zeta'\e^{\im\theta},\zeta'\e^{-\im\theta},\tau}\atop
	{\tau\zeta'\e^{\im\phi},\tau\zeta'\e^{-\im\phi}}}
\bigg|q,\tau\right].
\label{eq:sumHH}
\end{align}
To obtain a symmetric expression, we have used a transformation formula 
\begin{align}
{}_3\phi_2\left[
{{a,b,c}\atop{d,e}}\bigg| q,\frac{de}{abc}\right]
=\frac{(e/a,de/bc;q)_\infty}{(e,de/abc;q)_\infty}
{}_3\phi_2\left[
{{a,d/b,d/c}\atop{d,de/bc}}\bigg| q,e/a\right]
\end{align}
for generic $a,b,c,d,e$. 
Next, setting $t_1=\zeta'$, $t_2=s_1=s_2=0$ and $x=\e^{\im\theta}$, $y=\e^{-\im\theta}$ 
gives 
\begin{align}
&\sum_{n=0}^\infty \frac{\tau^n}{(q;q)_n}
H_n(\cos\theta;\zeta'|q)F_n(z,w;0)
\nonumber\\
&=\frac{(\tau\zeta'z,\tau\zeta'w;q)_\infty}
	{(\tau z\e^{\im\theta}, \tau z\e^{-\im\theta}, 
	\tau w\e^{\im\theta}, \tau w\e^{-\im\theta};q)_\infty}
{}_2\phi_2\left[
{{\zeta'\e^{\im\theta},\zeta'\e^{-\im\theta}}\atop
	{\tau\zeta'z,\tau\zeta'w}}
\bigg| q,\tau^2zw\right].
\label{eq:sumHF}
\end{align}
Here we have used Heine's transformation formula,
\begin{align}
{}_2\phi_1\left[
{{a,b}\atop{c}}\bigg| q,z\right]
=\frac{(az;q)_\infty}{(z;q)_\infty}
{}_2\phi_2\left[
{{a,c/b}\atop{c,az}}\bigg| q,bz\right].
\end{align}
With the help of the relation (\ref{eq:rec-bigHermite}), 
another kind of bilinear sum is obtained: 
\begin{align}
&\sum_{n=0}^\infty \frac{\tau^n}{(q;q)_n}
H_n(\cos\theta;\zeta'|q)H_{n+1}(\cos\phi;\zeta'|q)
\nonumber\\
&=\frac{1}{1-\tau^2}\Bigg[
\frac{2(\cos\phi-\tau\cos\theta)
		(\tau,\tau\zeta' \e^{\im\phi},\tau\zeta' \e^{-\im\phi};q)_\infty}
	{(\tau \e^{\im(\theta+\phi)},\tau \e^{\im(\theta-\phi)},
		\tau \e^{-\im(\theta+\phi)},\tau \e^{-\im(\theta-\phi)};q)_\infty}
{}_3\phi_2\left[
{{\zeta'\e^{\im\theta},\zeta'\e^{-\im\theta},\tau}\atop
	{\tau\zeta'\e^{\im\phi},\tau\zeta'\e^{-\im\phi}}}
\bigg|q,\tau\right]
\nonumber\\
&\qquad
-\frac{\zeta'(1-\tau)
		(\tau q,\tau\zeta' q\e^{\im\phi},\tau\zeta' q\e^{-\im\phi};q)_\infty}
	{(\tau q\e^{\im(\theta+\phi)},\tau q\e^{\im(\theta-\phi)},
		\tau q\e^{-\im(\theta+\phi)},\tau q\e^{-\im(\theta-\phi)};q)_\infty}
{}_3\phi_2\left[
{{\zeta'\e^{\im\theta},\zeta'\e^{-\im\theta},\tau q}\atop
	{\tau\zeta' q\e^{\im\phi},\tau\zeta' q\e^{-\im\phi}}}
\bigg|q,\tau q\right]
\Bigg].
\label{eq:sumHH2}
\end{align}
In order to take the limit $\tau\to 1$ in (\ref{eq:sumHH2}), 
we apply the following transformation formula for the first term of 
RHS of (\ref{eq:sumHH2}): 
\begin{align}
&{}_3\phi_2\left[
{{a,b,c}\atop{d,e}}\bigg| q,\frac{de}{abc}\right]
\nonumber\\
&=\frac{(e/b,e/c;q)_\infty}{(e,e/bc;q)_\infty}
{}_3\phi_2\left[
{{d/a,b,c}\atop{d,bcq/e}}\bigg| q,q\right]
+\frac{(d/a,b,c,de/bc;q)_\infty}{(d,e,bc/e,de/abc;q)_\infty}
{}_3\phi_2\left[
{{e/b,e/c,de/abc}\atop{de/bc,eq/bc}}\bigg| q,q\right].
\end{align}
As a result, we obtain
\begin{align}
&\sum_{n=0}^\infty 
\frac{H_n(\cos\theta;\zeta'|q)H_{n+1}(\cos\phi;\zeta'|q)}{(q;q)_n}
\nonumber\\
&=(\cos\phi-\cos\theta)
\left\{ (q,\zeta'\e^{\im\theta},\zeta'\e^{-\im\theta};q)_\infty
+(q,\zeta'\e^{\im\phi},\zeta'\e^{-\im\phi};q)_\infty\right\}
\bigg/ (\e^{\im(\theta+\phi)},\e^{\im(\theta-\phi)},
\e^{-\im(\theta+\phi)},\e^{-\im(\theta-\phi)};q)_\infty
\nonumber\\
&\qquad
-\zeta'\frac{(q,\zeta' q\e^{\im\theta},\zeta' q\e^{-\im\theta};q)_\infty}
	{(q\e^{\im(\theta+\phi)},q\e^{\im(\theta-\phi)},
	q\e^{-\im(\theta+\phi)},q\e^{-\im(\theta-\phi)};q)_\infty}
{}_3\phi_2\left[
{{q,\zeta' \e^{\im\phi},\zeta' \e^{-\im\phi}}\atop
{\zeta' q\e^{\im\theta},\zeta' q\e^{-\im\theta}}} \bigg| q,q\right].
\label{eq:sumHH3}
\end{align}

\section{Representation for the matrix products}
\label{sec:rep}

In the analysis of the steady state of the ASEP, 
full use of the matrix product method can be made if we have 
an explicit representation of the matrices and the vectors satisfying (\ref{eq:MPA}). 
In fact the matrices are closely related to the $q$-boson operators. 
If we set
\begin{align}
\D=\frac{1}{1-q}(1+\dd),\qquad
\E=\frac{1}{1-q}(1+\ee),\qquad
\A=\frac{1}{1-q}(\dd\ee-\ee\dd) ,
\label{eq:MPA2}
\end{align}
then $\dd$ and $\ee$ are the annihilation and creation operators of $q$-boson, 
respectively, and satisfy the relation
\begin{align}
\dd\ee-q\ee\dd=1-q.
\end{align}
On the basis of the $q$-deformed Fock states, 
the explicit representations of the matrices are as follows: 
\begin{subequations}
\begin{align}
&\dd=\left[
\begin{array}{cccc}
0 & \sqrt{1-q} & 0 & \cdots\\
0 & 0 & \sqrt{1-q^2} & \ddots\\
0 & 0 & 0 & \ddots\\
\vdots & \ddots & \ddots & \ddots
\end{array}
\right], \qquad
\ee=\left[
\begin{array}{cccc}
0 & 0 & 0 & \cdots\\
\sqrt{1-q} & 0 & 0 & \ddots\\
0 & \sqrt{1-q^2} & 0 & \ddots\\
\vdots & \ddots & \ddots & \ddots
\end{array}
\right],
\end{align}
\begin{align}
&\A=\mathrm{diag}(1,q,q^2,\cdots).
\end{align}
\end{subequations}
The boundary vectors are determined from (\ref{eq:MPA-W}) and (\ref{eq:MPA-V}), yielding
\begin{align}
\< W|=(w_0, w_1,\cdots)^T,\qquad
|V\> =(v_0, v_1,\cdots),
\end{align}
where
\begin{subequations}
\begin{align}
&w_n=F_n(a,c;0)\Big/\sqrt{(q;q)_n},\\
&v_n=F_n(b,d;0)\Big/\sqrt{(q;q)_n}
\end{align}
\end{subequations}
Here we have introduced more useful boundary parameters: 
\begin{align}
a=\kappa_{\alpha,\gamma}^+,\qquad b=\kappa_{\beta,\delta}^+,\qquad
c=\kappa_{\alpha,\gamma}^-,\qquad d=\kappa_{\beta,\delta}^-,
\end{align}
where 
\begin{align}
\kappa_{x,y}^\pm=
\frac{1}{2x}\left[ (1-q-x+y)\pm\sqrt{(1-q-x+y)^2+4xy}\right].
\end{align}
For $\alpha, \beta, \gamma, \delta$ positive, 
the parameters satisfy the inequalities $a>0, b>0, -1<c<0, -1<d<0$. 

Set $\zeta'=(1-q)\zeta$. 
We introduce a vector, called the bulk vector, 
\begin{align}
|h_{\zeta'}(\cos\theta)\>=
(h_{0}(\cos\theta;\zeta'),h_{1}(\cos\theta;\zeta'),h_{2}(\cos\theta;\zeta'),\cdots)^T,
\end{align}
where the $n$th element is 
\begin{align}
h_n(\cos\theta;\zeta')=H_n(\cos\theta;\zeta'|q)\Big/\sqrt{(q;q)_n}.
\end{align}
From the three-term recurrence relation of the continuous big $q$-Hermite polynomial 
(\ref{eq:rec-bigHermite}), we have the eigenvalue equation 
\begin{align}
(\dd+\zeta'\A+\ee)|h_{\zeta'}(x)\>=2x|h_{\zeta'}(x)\>.
\label{eq:eigen-h}
\end{align}
In the terminology of the orthogonal polynomial, 
the matrix $(\dd+\zeta'\A+\ee)$ is the 
Jacobi operator for the continuous big $q$-Hermite polynomial. 
We define $\< h_{\zeta'}(\cos\theta)|$ as the transpose of $|h_{\zeta'}(\cos\theta)\>$.
The orthogonality relation (\ref{eq:ortho-bigHermite}) guarantees 
the completeness of $|h_{\zeta'}(x)\>$: 
\begin{align}
1=\int_{0}^\pi \frac{\d\theta}{2\pi}
\frac{(q,\e^{2\im\theta},\e^{-2\im\theta};q)_\infty}
	{(\zeta' \e^{\im\theta},\zeta' \e^{-\im\theta};q)_\infty}
|h_{\zeta'}(\cos\theta)\>\< h_{\zeta'}(\cos\theta)|.
\label{eq:comp-h}
\end{align}
Equation (\ref{eq:comp-h}) shows that 
the spectrum of $(\dd+\zeta'\A+\ee)$ is $\{2\cos\theta ;\theta\in[0,\pi]\}= [-2,2]$. 

In fact, the representation given here is only valid when $|a|,|b|,|c|,|d|<1$; 
otherwise, the inner products of the boundary vectors and the bulk vectors 
would become divergent series and ill-defined. 
Therefore, the quantities of the ASEP should be calculated first in the 
well-defined region of 
the boundary parameters and then analytically continued to other regions. 
An explicit representation available for all parameters is not found yet.

\section{Nonequilibrium phase transitions}
\label{sec:analysis}

Using the explicit representation in the previous section, 
we derive integral formulae for several quantities of the two-species ASEP. 
Evaluation of the integrals is done in the thermodynamic limit $L,N \gg 1$. 

We begin with the partition function. 
First, we set $\xi=1$ and assume $|a|,|b|,|c|,|d|<1$. 
By the use of (\ref{eq:MPA2}), (\ref{eq:eigen-h}) and (\ref{eq:comp-h}), we have
\begin{align}
&Z_L(\zeta)\nonumber\\
&=\< W|(2+\dd +\zeta'\A+\ee)^L|V\> (1-q)^{-L}\nonumber\\
&=\int_{0}^\pi \frac{\d\theta}{2\pi}
\frac{(q,\e^{2\im\theta},\e^{-2\im\theta};q)_\infty}
	{(\zeta' \e^{\im\theta},\zeta' \e^{-\im\theta};q)_\infty}
\< W|(2+\dd +\zeta'\A+\ee)^L|h_{\zeta'}(\cos\theta)\>
\< h_{\zeta'}(\cos\theta)|V\> (1-q)^{-L}\nonumber\\
&=\int_{0}^\pi \frac{\d\theta}{2\pi}
\frac{(q,\e^{2\im\theta},\e^{-2\im\theta};q)_\infty}
	{(\zeta' \e^{\im\theta},\zeta' \e^{-\im\theta};q)_\infty}
\< W|h_{\zeta'}(\cos\theta)\>(2+2\cos\theta)^L
\< h_{\zeta'}(\cos\theta)|V\> (1-q)^{-L}.
\end{align}
The inner products in the integral are calculated from (\ref{eq:sumHF}). 
In this way, we arrive at the following integral expression:
\begin{align}
Z_L(\zeta)=
\int_{0}^\pi \frac{\d\theta}{2\pi}&
\frac{(q,\zeta' a,\zeta' b,\zeta' c,\zeta' d,\e^{2\im\theta},\e^{-2\im\theta};q)_\infty}
{(\zeta' \e^{\im\theta},\zeta'\e^{-\im\theta},
a\e^{\im\theta},a\e^{-\im\theta},
b\e^{\im\theta},b\e^{-\im\theta},
c\e^{\im\theta},c\e^{-\im\theta},
d\e^{\im\theta},d\e^{-\im\theta};q)_\infty} \nonumber\\
&\times
{}_2\phi_2\left[ {{\zeta' \e^{\im\theta},\zeta'\e^{-\im\theta}}\atop{\zeta' a,\zeta' c}}
\Bigg| q,ac\right]
{}_2\phi_2\left[ {{\zeta' \e^{\im\theta},\zeta'\e^{-\im\theta}}\atop{\zeta' b,\zeta' d}}
\Bigg| q,bd\right]
\nonumber\\
&\times
[2(1+\cos\theta)]^L (1-q)^{-L}.
\end{align}
Rewriting this expression with $z$ s.t. $z+z^{-1}=2\cos\theta$ yields 
\begin{align}
Z_L(\zeta)=
\oint \frac{\d z}{4\pi \im z}&
\frac{(q,\zeta' a,\zeta' b,\zeta' c,\zeta' d,z^2,z^{-2};q)_\infty}
{(\zeta' z,\zeta'/z,az,a/z,bz,b/z,cz,c/z,dz,d/z;q)_\infty} \nonumber\\
&\times
{}_2\phi_2\left[ {{\zeta' z,\zeta'/z}\atop{\zeta' a,\zeta' c}}\Bigg| q,ac\right]
{}_2\phi_2\left[ {{\zeta' z,\zeta'/z}\atop{\zeta' b,\zeta' d}}\Bigg| q,bd\right]
\nonumber\\
&\times
[(1+z)(1+z^{-1})]^L (1-q)^{-L}.
\label{eq:Z_L-integral0}
\end{align}
Here the integral is along the unit circle. 
Now, we consider other regions of the boundary parameters. 
Since the integrand has a number of poles that move depending on the 
boundary parameters, analytic continuation of the integral is made 
by deforming the integral contour 
such that no pole crosses the contour while changing the parameters. 
Therefore, the contour should always enclose the poles $\epsilon q^k$ 
and always exclude the poles  $(\epsilon q^k)^{-1}$ for $k\in \mathbb{Z}_{\ge 0}$ and
$\epsilon=a,b,c,d,\zeta'$. 
Let $M_\epsilon=\mathrm{max}\big\{m\in \mathbb{Z}_{\ge 0}; \epsilon q^m>1\big\}$. 
If we suppose $f(z)$ has no pole at $z=\epsilon q^m, (\epsilon q^m)^{-1}$, 
$m=0,\cdots,M_\epsilon$ for 
$\epsilon=a,b,c,d,\zeta'$, then the contour integral can be decomposed into 
two parts; the sum of the residues at $z=\epsilon q^m, (\epsilon q^m)^{-1}$, 
$m=0,\cdots,M_\epsilon$ and an integral along the unit circle, 
\begin{align}
&\oint\frac{\d z}{4\pi \im z}
\frac{(z^2,z^{-2};q)_\infty}
{(\zeta' z,\zeta'/z,az,a/z,bz,b/z,cz,c/z,dz,d/z;q)_\infty} 
f(z)\nonumber\\
&=\sum_{m=0}^{M_{a}}\frac{(a^2q^{2m},a^{-2}q^{-2m};q)_\infty}
{(q^{-m};q)_m (q,a^2q^{m},\zeta' aq^m,\zeta'/a q^{-m},
	abq^m,b/a q^{-m},acq^m,c/a q^{-m},adq^m,d/a q^{-m};q)_\infty}
f(aq^m)
\nonumber\\
&\qquad+\big[ a\leftrightarrow b,c,d,\zeta'\big]\nonumber\\
&\qquad+\oint_{\mathrm{unit\ circle}}\frac{\d z}{4\pi \im z}
\frac{(z^2,z^{-2};q)_\infty}
{(\zeta' z,\zeta'/z,az,a/z,bz,b/z,cz,c/z,dz,d/z;q)_\infty} 
f(z).
\label{eq:integral-sum}
\end{align}
With this decomposition, the integral expression (\ref{eq:Z_L-integral0}) is 
valid  for all parameter regions. 
Introducing the fugacity for the particle $2$ with $\xi^2$ requires the 
replacement 
$a\to \xi^{-1}a, b\to\xi b, c\to\xi^{-1}c, d\to\xi d, \zeta'\to	\zeta'\xi^{-1}$. 
The grand canonical partition function is therefore obtained as 
\begin{align}
Z_L(\xi^2,\zeta)=
\oint \frac{\d z}{4\pi \im z}&
\frac{(q,\zeta' \xi^{-2}a,\zeta' b,\zeta' \xi^{-2}c,\zeta' d,z^2,z^{-2};q)_\infty}
{(\zeta'\xi^{-1} z,\zeta'\xi^{-1}/z,\xi^{-1}az,\xi^{-1}a/z,
\xi bz,\xi b/z,\xi^{-1}cz,\xi^{-1}c/z,\xi dz,\xi d/z;q)_\infty} \nonumber\\
&\times
{}_2\phi_2\left[ {{\zeta'\xi^{-1} z,\zeta'\xi^{-1}/z}\atop
{\zeta' \xi^{-2}a,\zeta'\xi^{-2} c}}
\Bigg| q,\xi^{-2}ac\right]
{}_2\phi_2\left[ {{\zeta'\xi^{-1} z,\zeta'\xi^{-1}/z}\atop{\zeta' b,\zeta' d}}
\Bigg| q,\xi^2 bd\right]
\nonumber\\
&\times
[(1+\xi z)(1+\xi z^{-1})]^L (1-q)^{-L}.
\label{eq:Z_L-integral}
\end{align}
We can see that $\zeta=0$ reproduces the one-species partition function 
in \cite{USW} up to a factor. 

Let us examine the thermodynamic limit, large $L$ and $N=L\rho_1$. 
It is difficult to evaluate the integral (\ref{eq:Z_L-integral}), 
but the asymptotic behavior for large $L$ is obtained relatively easily. 
We should make a careful remark. 
The relation between the fugacity $\zeta$ and the particle density $\rho_1$ is 
given by (\ref{eq:rho_1}). 
Unfortunately, the one-to-one correspondence is not valid for all parameters. 
Thus, we should also switch to the canonical ensemble for some cases. 

First, we suppose that $\zeta'>a,b$ and $\zeta'>1$. 
Recall that $\zeta'=(1-q)\zeta$. 
The decomposition (\ref{eq:integral-sum}) shows that 
the leading term is from the pole $z=\zeta', 1/\zeta'$ and of order 
$[(1+\zeta')(1+\xi^2/\zeta')]^L$. 
In this case, we have $\rho_1=\frac{\zeta'-1}{\zeta'+1}$ from (\ref{eq:rho_1}). 
Note that $\rho_1$ is a positive increasing function of $\zeta'$ and satisfies 
$\rho_1\to 0$ as $\zeta'\to1$ and $\rho_1\to 1$ as $\zeta'\to\infty$. 
Hence, the one-to-one correspondence is valid in this parameter region. 
If we put $\zeta_0\equiv\frac{1+\rho_1}{1-\rho_1}$, 
the condition for the one-to-one correspondence is equivalent to $\zeta_0>a,b$. 
The canonical partition function is just obtained by the relation 
\begin{align}
Z_L(\xi^2,\zeta)\simeq Z_{L,N}(\xi^2)\zeta^N.
\end{align}

For other parameter regions, the one-to-one correspondence of $\rho_1$ and $\zeta$ 
fails since the $N=0$ part of the grand canonical partition function becomes 
dominant and the $N\neq 0$ part does not appear in its asymptotic form. 
Now, suppose that $a>\zeta_0$, $a>b$. 
In the case where the number of the particle $1$ is fixed to $N$, the partial sum  is 
expanded as 
\begin{align}
Z_{L,N}(\xi^2)&=\oint\frac{\d\zeta}{2\pi \im\zeta^{N+1}} Z_L(\xi^2,\zeta)
\nonumber\\
&=\sum_{\ell=0}^\infty \mathop{\mathrm{Res}}_{\zeta'=(\xi^{-1}zq^\ell)^{-1}}
{\zeta^{-(N+1)}} Z_L(\xi^2,\zeta)d\zeta
+\sum_{\ell=0}^\infty \mathop{\mathrm{Res}}_{\zeta'=(\xi^{-1}/zq^\ell)^{-1}}
{\zeta^{-(N+1)}} Z_L(\xi^2,\zeta)d\zeta
\nonumber\\
&=\sum_{\ell=0}^\infty \oint\frac{\d z}{4\pi \im z}
-\left(\xi^{-1}z(1-q)q^\ell\right)^N
\frac{(\xi^{-1}a/zq^{-\ell},\xi b/zq^{-\ell},
	\xi^{-1}c/zq^{-\ell},\xi d/zq^{-\ell};q)_\ell
	(z^2;q)_\infty}
{(q^{-\ell},q^{-\ell}/z^2;q)_\ell (\xi^{-1}azq^{-\ell},\xi bzq^{-\ell},
	\xi^{-1}czq^{-\ell},\xi dzq^{-\ell};q)_\infty}
\nonumber\\
&\qquad\qquad\qquad\times
{}_2\phi_2\left[
{{q^{-\ell},q^{-\ell}/z^2}\atop{\xi^{-1}a/zq^{-\ell},\xi^{-1}c/zq^{-\ell}}}
\bigg| q,\xi^{-2}ac\right]
{}_2\phi_2\left[
{{q^{-\ell},q^{-\ell}/z^2}\atop{\xi b/zq^{-\ell},\xi c/zq^{-\ell}}}
\bigg| q,\xi^{2}bd\right]
\nonumber\\
&\qquad\qquad\qquad\times
[(1+\xi z)(1+\xi/z)]^L (1-q)^{-L}
\nonumber\\
&\qquad +[z\to 1/z].
\label{eq:keisan}
\end{align}
Integrating with respect to $z$ with the decomposition (\ref{eq:integral-sum}) 
gives the leading term 
from the pole $z=(\xi^{-1}aq^{-\ell})^{-1}$ for a small $\ell$. 
Note that the second term of (\ref{eq:keisan}) 
gives the same value as the first term. 
Due to the factor $q^{\ell N}$, the leading term of $Z_{L,N}$ is the term 
taking both $\ell=0$ in the summation and 
the residue at the pole $z=(\xi^{-1}a)^{-1}$. 
The same argument holds for the case where $b>\zeta_0$, $b>a$. 

With these evaluations of the partition functions, 
the particle current and the particle densities are calculated from 
(\ref{eq:rho_2})--(\ref{eq:J_2-2}). 
As a whole, we have the following behavior in each parameter region: 

(A) $a>\zeta_0$, $a>b$
\begin{align}
Z_{L,N}(\xi^2)\simeq Z_{L,N}^{(a)}(\xi^2)
=-\frac{(\xi^2a^{-2};q)_\infty(1-q)^{N-L}}
	{(q,\xi^2 b/a,c/a,\xi^2 d/a;q)_\infty}
	a^{-N}[(1+a)(1+\xi^2 a)]^L,
\label{eq:phaseA}
\end{align}

\begin{align}
\rho_2=\frac{1}{1+a},\qquad
\Delta\rho_2^2=\frac{a}{(1+a)^2L},\qquad
J_2=(1-q)\frac{a}{(1+a)^2}.
\end{align}

(B) $b>\zeta_0$, $b>a$
\begin{align}
Z_{L,N}(\xi^2)\simeq Z_{L,N}^{(b)}(\xi^2)
=-\frac{(\xi^{-2}b^{-2};q)_\infty(1-q)^{N-L}}
	{(q,\xi^{-2}a/b,\xi^{-2}c/b,d/b;q)_\infty}
	(\xi^2 b)^{-N}[(1+\xi^2 b)(1+1/b)]^L,
\label{eq:phaseB}
\end{align}

\begin{align}
\rho_2=\frac{b}{1+b}-\rho_1,\qquad
\Delta\rho_2^2=\frac{b}{(1+b)^2L},\qquad
J_2=(1-q)\frac{b}{(1+b)^2}.
\end{align}

(C) $a, b<\zeta_0$
\begin{align}
Z_{L,N}(\xi^2)\simeq Z_{L,N}^{(\zeta_0)}(\xi^2)
=-\frac{(\zeta_0^{-2}\xi^2;q)_\infty (1-q)^{N-L}}
	{(a/\zeta_0,\xi^2 b/\zeta_0, c/\zeta_0,\xi^2 d/\zeta_0;q)_\infty}
	\zeta_0^{-N}[(1+\zeta_0)(1+\xi^2/\zeta_0)]^L,
\label{eq:phaseC}
\end{align}

\begin{align}
\rho_2=\frac{1-\rho_1}{2},\qquad
\Delta\rho_2^2=\frac{1-\rho_1^2}{4L},\qquad
J_2=(1-q)\frac{1-\rho_1^2}{4}.
\end{align}
It is interesting to observe that 
\begin{align}
Z_{L,N}^{(a,b)}(\xi^2)\simeq
\oint\frac{\d\zeta_0}{2\pi \im\zeta_0^{N+1}} Z_{L,N}^{(\zeta_0)}(\xi^2).
\end{align}

\begin{figure}[tbp]
\begin{center}
\includegraphics[scale=0.5]{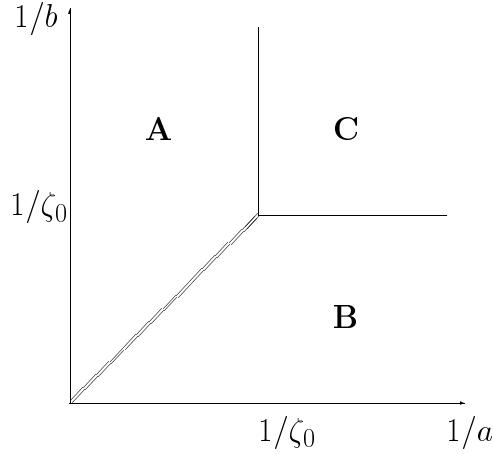}
\end{center}
\caption{The phase diagram of the two-species ASEP. 
There are three phases; 
the phase A: the low-density phase, the phase B: the high-density phase 
and the phase C: the maximal-current phase. 
The double line indicates the transition of the first order 
and the solid line indicates the transition of the second order.}
\label{fig:pd}
\end{figure}

With respect to the behavior of the particle 2,
we have three phases; 
just as the one-species ASEP with open boundaries, 
which is recovered by putting $\rho_1=0$ in the two-species model. 
Accordingly, the three phases are 
the phase A: the low-density phase, the phase B: the high-density phase 
and the phase C: the maximal current phase. See Figure \ref{fig:pd}. 
By checking the discontinuity of the derivatives of the partition function, 
the phase transition is found to be 
of the second order between the phases A(B) and C and 
of the first order between the phases A and B. 
On the critical line between the phases A and B, 
the low-density phase and the high-density phase coexist. 
Indeed, a shock profile of the density of the particle $2$ 
is expected on this coexisting line. 
In the phase A, the density of the particle $2$ 
is not suppressed by the existence of the particle $1$. 
On the other hand, 
in the phase B, 
the density of the particle $2$ is suppressed by $\rho_1$, and 
in the phase C, suppressed by $\rho_1/2$. 
The existence of the particle $1$ also causes the suppression of the maximal value 
of the current of the particle $2$ from $(1-q)/4$ to $(1-q)(1-\rho_1^2)/4$ 
and hence shifts the lines of phase transitions between the phases A and C, and 
between the phases B and C. 

The physical explanation of the shift of the critical line 
between the phases A(B) and C is as follows. 
As mentioned before, the dynamics of the particle 2 is unaffected 
by the presence of the particle 1. 
However, due to the exclusion interaction between different species, 
the rates of the injection and removal of the particle 2 at the boundaries 
have effective values, depending on the density of the particle 1 
near the boundaries. 
The phase diagram may be the same as the one for the one-species, 
drawn with respect to the effective boundary parameters. 
Instead, with respect to the original parameters, 
it appears that the critical lines are shifted.

We can explicitly obtain the $n$-point function in an integral form. 
Similarly to the partition function, 
we insert the completeness relation (\ref{eq:comp-h}) 
$n+1$ times in the matrix products, 
and arrive at an integral formula. 
Let $\eta_i=\tau_i$ or $\sigma_i$. The result is as follows: 
\begin{align}
\<\eta_{i_1}\cdots\eta_{i_n}\>=&
\frac{1}{Z_L}\frac{1}{(1-q)^L}
\prod_{\ell=1}^{n+1}\left[
\oint_{C_\ell}\frac{\d z_\ell}{4\pi \im z_\ell}
\frac{(q,z_\ell^2,z_\ell^{-2};q)_\infty}{(\zeta'z_\ell,\zeta'/z_\ell;q)_\infty}
[(1+z_\ell)(1+z_\ell^{-1})]^{i_\ell-i_{\ell-1}-1}
\right]
\nonumber\\
&\times
{}_2\phi_2\left[ {{\zeta' z_1,\zeta'/z_1}\atop{\zeta' a,\zeta' c}}
\Bigg| q,ac\right]
{}_2\phi_2\left[ {{\zeta' z_{n+1},\zeta'/z_{n+1}}\atop{\zeta' b,\zeta' d}}
\Bigg| q,bd\right]
\nonumber\\
&\times
(az_1,a/z_1,cz_1,c/z_1,bz_{n+1},b/z_{n+1},dz_{n+1},d/z_{n+1};q)_\infty^{-1}
\nonumber\\
&\times
\prod_{k=1}^n \left[\chi_{\eta_{i_k}=\tau_{i_k}} \mathcal{T}(z_k,z_{k+1})+
\chi_{\eta_{i_k}=\sigma_{i_k}} \mathcal{S}(z_k,z_{k+1})\right],
\end{align}
where $i_0=0$ and $i_{n+1}=L+1$ in the products, and
$\chi$ denotes the characterstic function: 
$\chi_{\mathrm{TRUE}}=1, \chi_{\mathrm{FALSE}}=0$. 
The contour $C_\ell$ ($\ell=2,\cdots,n+1$) encloses the poles at 
$z_\ell=z_{\ell-1}q^k,z_{\ell-1}^{-1}q^k,\zeta' q^k$ and
excludes the poles at 
$z_\ell=(z_{\ell-1}q^k)^{-1},(z_{\ell-1}^{-1}q^k)^{-1},(\zeta' q^k)^{-1}$ 
($k\in\mathbb{Z}_{\ge 0}$).
The contour $C_1$ encloses the poles at 
$z_1=aq^k,cq^k$ 
and excludes the poles at 
$z_1=(aq^k)^{-1}, (cq^k)^{-1}$ ($k\in\mathbb{Z}_{\ge 0}$). 
The contour $C_{n+1}$ also encloses the poles at 
$z_{n+1}=bq^k, dq^k$ 
and also excludes the poles at
$z_{n+1}=(bq^k)^{-1}, (dq^k)^{-1}$ ($k\in\mathbb{Z}_{\ge 0}$). 
By the use of the summation formula (\ref{eq:sumHH3}), 
$\<h(\cos\theta_k)|(1+\dd)|h(\cos\theta_{k+1})\>$ is summed up as 
\begin{align}
\mathcal{T}(z_k,z_{k+1})=
&	4\pi \im z_{k+1}\frac{(\zeta'z_{k+1},\zeta'/z_{k+1};q)_\infty}
		{(q,z_{k+1}^2,z_{k+1}^{-2};q)_\infty} \delta(z_{k+1}-z_k)
		\nonumber\\
&	+\frac{1}{2}(z_{k+1}+1/z_{k+1}-z_k-1/z_k)\left\{
	(q,\zeta'z_k,\zeta'/z_k;q)_\infty+
	(q,\zeta'z_{k+1},\zeta'/z_{k+1};q)_\infty \right\}
	\nonumber\\
&	\qquad\times
	(z_kz_{k+1},z_k/z_{k+1},z_{k+1}/z_k,1/z_kz_{k+1};q)_\infty^{-1}
	\nonumber\\
&	-\frac{\zeta'}{2}\frac{(q,q\zeta'z_k,q\zeta'/z_k;q)_\infty}
		{(qz_kz_{k+1},qz_k/z_{k+1},qz_{k+1}/z_k,q/z_kz_{k+1};q)_\infty}
	{}_3\phi_2\left[
	{{q,\zeta'z_{k+1},\zeta'/z_{k+1}}\atop{q\zeta'z_k,q\zeta'z_k}}
	\bigg| q,q\right].
\end{align}
By the use of the formula (\ref{eq:sumHH}) with $\tau=q$, 
$\<h(\cos\theta_k)|\zeta' \A|h(\cos\theta_{k+1})\>$ is calculated as 
\begin{align}
\mathcal{S}(z_k,z_{k+1})=
	\frac{\zeta'(q,q\zeta'z_k,q\zeta'/z_k;q)_\infty}
		{(qz_kz_{k+1},qz_k/z_{k+1},qz_{k+1}/z_k,q/z_kz_{k+1};q)_\infty}
	{}_3\phi_2\left[
	{{q,\zeta'z_{k+1},\zeta'/z_{k+1}}\atop{q\zeta'z_k,q\zeta'z_k}}
	\bigg| q,q\right].
\end{align}
It is difficult to evaluate the $n$-point functions. 
The phase transitions and the density profiles 
for the two-species ASEP with specialization 
$q=\gamma=\delta=0$ 
were discussed in \cite{Arita,Arita06}. 
In fact, in the phase A and the phase B, 
phase separation occurs. 
An intuitive explanation is that 
if there are many particles $2$, by the exchange rule 
$21\ \mathop{\rightleftarrows}^1_q\ 12$, 
the particles $1$ are pushed from right to left 
and are accumulated near the left boundary. 
By the particle-hole symmetry, 
if there are a few particles $2$, the particles $1$ are 
accumulated near the right boundary.

\section{Conclusion}

We have studied the two-species Asymmetric Simple Exclusion Process (ASEP) with 
open boundaries. For the steady state, we have calculated the partition function 
through the matrix product method and obtained integral formulae. 
We examine the asymptotic behavior of the integral in the thermodynamic limit. 
Accordingly, we have found three phases depending on the boundary parameters, 
just as the one-species version of the ASEP with open boundaries. 
We have furthermore calculated the $n$-point functions and obtained integral formulae. 
In the analysis, the $q$-calculus is heavily exploited and the continuous 
big $q$-Hermite polynomials is important. 
To extend the relation between nonequilibrium particle models and 
the theory of $q$-orthogonal polynomials may be a fascinating future problem.

\section*{Acknowledgement}

The author would like to express his sincere gratitude to 
Professor Miki Wadati for continuous encouragements.

\end{document}